\documentclass[journal]{IEEEtran}
\usepackage{graphicx,color}
\usepackage{cite}
\usepackage{setspace} 
\usepackage{amsmath}
\usepackage{amsmath,amsthm,amssymb}
\usepackage{multirow}
\usepackage{hhli ne}
\usepackage{epsfig}
\usepackage{subcaption}
\usepackage{epstopdf}
\usepackage{verbatim}
\usepackage{algorithm}
\usepackage{algorithmicx}
\usepackage{algpseudocode}
\usepackage{etoolbox}
\usepackage{cases}
\usepackage{csquotes}
\usepackage{enumitem}
\usepackage{mathtools,stmaryrd}
\SetSymbolFont{stmry}{bold}{U}{stmry}{m}{n}
\usepackage{bbm}
\usepackage{lettrine}
\usepackage{nicematrix}
\usepackage{steinmetz}

\newcommand{\suchthat}{\;\ifnum\currentgrouptype=16 \middle\fi|\;}

\makeatletter
\newcommand*{\indep}{%
  \mathbin{%
    \mathpalette{\@indep}{}%
  }%
}
\newcommand*{\nindep}{%
  \mathbin{
    \mathpalette{\@indep}{\not}
  }%
}
\newcommand*{\@indep}[2]{%
  \sbox0{$#1\perp\m@th$}
  \sbox2{$#1=$}
  \sbox4{$#1\vcenter{}$}
  \rlap{\copy0}
  \dimen@=\dimexpr\ht2-\ht4-.2pt\relax
  \kern\dimen@
  {#2}%
  \kern\dimen@
  \copy0 
} 
\makeatother

\makeatletter
\newcommand*{\algrule}[1][\algorithmicindent]{%
  \makebox[#1][l]{%
    \hspace*{.2em}
    \vrule height .75\baselineskip depth .25\baselineskip
  }
}

\newcount\ALG@printindent@tempcnta
\def\ALG@printindent{%
    \ifnum \theALG@nested>0
    \ifx\ALG@text\ALG@x@notext
    \else
    \unskip
    \ALG@printindent@tempcnta=1
    \loop
    \algrule[\csname ALG@ind@\the\ALG@printindent@tempcnta\endcsname]%
    \advance \ALG@printindent@tempcnta 1
    \ifnum \ALG@printindent@tempcnta<\numexpr\theALG@nested+1\relax
    \repeat
    \fi
    \fi
}
\patchcmd{\ALG@doentity}{\noindent\hskip\ALG@tlm}{\ALG@printindent}{}{\errmessage{failed to patch}}
\patchcmd{\ALG@doentity}{\item[]\nointerlineskip}{}{}{} 
\makeatother

\newtheorem{prop}{Proposition}

\allowdisplaybreaks

\begin{document}

\title{Chaotic Noncoherent SWIPT in\\
Multi-Functional RIS-Aided Systems \vspace{-2mm}}

\author{Authors}
\author{Priyadarshi Mukherjee, \textit{Senior Member, IEEE}, Constantinos Psomas, \textit{Senior Member, IEEE},\\
Himal A. Suraweera, \textit{Senior Member, IEEE}, and Ioannis Krikidis, \textit{Fellow, IEEE}
\thanks{P. Mukherjee is with the Department of Electrical Engineering and Computer Science, Indian Institute of Science Education and Research  Bhopal, India (e-mail: priyadarshi@ieee.org). C. Psomas is with the Department of Computer Science and Engineering, European University  Cyprus, Cyprus (e-mail: c.psomas@euc.ac.cy). H. A. Suraweera is with the Department of Electrical and Electronic Engineering, University of Peradeniya, Peradeniya, Sri Lanka (e-mail: himal@eng.pdn.ac.lk). I. Krikidis is with the Department of Electrical and Computer Engineering, University of Cyprus, Cyprus (e-mail: krikidis@ucy.ac.cy).}\vspace{-6.8mm}}

\maketitle

\begin{abstract}
In this letter, we investigate the design of chaotic signal-based transmit
waveforms in a multi-functional reconfigurable intelligent surface (MF-RIS)-aided set-up for simultaneous wireless information and power transfer. We propose a differential chaos shift keying-based MF-RIS-aided set-up, where the MF-RIS is partitioned into three non-overlapping surfaces. The elements of the first sub-surface perform energy harvesting (EH), which in
turn, provide the required power to the other two sub-surfaces responsible for transmission and reflection of the incident signal. By considering a frequency selective scenario and a realistic EH model, we characterize the chaotic MF-RIS-aided system in terms of its EH performance and the associated bit error rate. Thereafter, we characterize the harvested energy-bit error rate trade-off and derive a lower bound on the number of elements required to operate in the EH mode. Accordingly, we propose novel transmit waveform designs to demonstrate the importance of the choice of appropriate system parameters in the context of achieving self-sustainability.
\end{abstract}


\IEEEpeerreviewmaketitle

\section{Introduction}
\lettrine[lines=2]{W}{ith} sustainability and scalability coming out as two significant challenges for implementing the fifth generation (5G) and beyond wireless technology, worldwide research initiatives are increasingly exploring innovative solutions such as simultaneous wireless information and power transfer (SWIPT) \cite{psomas6g}. The authors in \cite{chaosexp1} experimentally demonstrate that signals with high peak-to-average-power-ratio, such as chaotic signals, provide higher wireless power transfer (WPT) efficiency. Moreover, the work in \cite{wcl2} proposes an analytical framework for chaotic WPT, which supports the above observation. Furthermore, the authors in \cite{jstsp} investigate a noncoherent differential chaos shift keying (DCSK)-based framework and propose WPT optimal transmit waveform designs. Hence, to exploit this additional advantage for both information and power transfer, we choose DCSK-based waveforms to explore its implications for SWIPT. The study in \cite{chaoswipt4} proposes various link selection schemes in a relay-based DCSK-SWIPT network and the authors in \cite{twc} investigate a DCSK-based multi-antenna receiver architecture for SWIPT.

Moreover, to accommodate the rising demand of future networks with flexible and reconfigurable characteristics, a new technology termed as reconfigurable intelligent surface (RIS) has been recently proposed \cite{r1r1,r12minor}. A self-sustainable RIS \cite{zeris} controls the propagation environment via software-controlled metasurfaces and its energy requirement is met from the incident signal via the process of energy harvesting (EH). The authors in \cite{tcom} investigate the problem of transmit waveform design in a self-sustainable RIS-assisted DCSK-based set-up. This work focuses on the reflection-only RISs, i.e., this RIS is capable of only reflecting the incident signal. Both the transmitter (Tx) and the receiver (Rx) need to be located on the same side of the RIS. In this context, the work in \cite{sris} proposes the novel concept of a simultaneously transmitting and reflecting (STAR) RIS, capable of $360^\circ$ radio environment  implementation. But still, the RIS transmitted/reflected signal is attenuated twice, which severely attenuates signal reaching the Rx. As a solution, the work in \cite{mfris1} proposes a new multi-functional (MF) RIS, which comprises of an array of active elements. Each array element exhibits a three-layer structure, which enables multidimensional and full-space signal manipulation. An MF-RIS essentially harvests energy from the incident signal, which in turn, is used for simultaneous transmission and reflection of the amplified incident signals.

Accordingly, by considering a frequency selective  scenario and a nonlinear EH model, we propose a novel DCSK-based MF-RIS-aided noncoherent SWIPT architecture. Specifically, the MF-RIS is partitioned into three non-overlapping surfaces, with one acting in the EH mode, and the other two operating in the information transmission (IT) and information reflection (IR) mode, respectively. We derive analytical expression for the system's bit error rate (BER) and a lower bound on the number of elements required to operate in the EH mode. We investigate the BER-harvested power trade-off and propose appropriate transmit waveform designs for achieving self-sustainability. Results demonstrate the impact of this trade-off in the context of appropriate choice of system parameters.

\section{System Model}
We consider a set-up with a single antenna Tx and two Rxs, each equipped with a single antenna. With no direct link between the Tx and both the Rxs, an MF-RIS, consisting of $N$ elements, is employed to realize the entire communication process. Since the MF-RIS is able to support complete $360^{\circ}$ coverage, we assume that the two Rxs, namely, $U_t$ and $U_r$, to be located in each of its transmission and reflection space, respectively. Also, we consider the adjacent elements of the MF-RIS to have at least half-wavelength spacing, i.e., the associated wireless channels are independent.

\subsection{SR-DCSK Signals}  \label{dcskdef}
In conventional DCSK, each information bit is characterized by two sets of equal length chaotic samples; the first set represents the reference and the other conveys information. On the contrary, the short reference DCSK (SR-DCSK) symbol has a flexible and shorter reference length, i.e., the SR-DCSK-based  Tx output during the $p$-th transmission interval is \cite{jstsp}
\begin{align}  \label{symsr}
& s_{p,q} \nonumber\\
 &=\!\!\begin{cases} 
\!\! x_{p,q}, & \!\!\! q\!=\!(p-1)(\beta+\phi)+1,\dots,(p-1)\beta+p\phi,\\
\!\! d_px_{p,q-\phi}, & \!\!\!q\!=\!(p-1)\beta+p\phi+1,\dots,p(\beta+\phi),
\end{cases}&
\end{align}
where $d_p=\pm1$ is the information symbol, $x_{p,q}$ is the chaotic reference, and $x_{p,q-\phi}$ is its delayed version. Here, $\beta \in \mathbb{Z}^+$ is the spreading factor and $x_{p,q}$ is generated according to various existing chaotic maps. In this work, we consider the Chebyshev map $x_{q+1}=1-2x_q^2$ for the generation of chaotic sequences. The reason for this choice  is its good auto/cross correlation properties, i.e., chaotic signals generated with different initial values to be regarded as quasi-orthogonal. The chaotic component of length $\phi$ is followed by $\zeta$ repetitions of its data modulated replica, such that $\beta=\zeta\phi$.

\subsection{MF-RIS Characterization}  \label{srischarc}
The elements of the MF-RIS are grouped into three parts, namely EH, IT, and IR section, consisting of $N_h,N_t,$ and $N_r$ elements, respectively, with $N_h+N_t+N_r=N$ \cite{mfris1}. The EH section completely absorbs the energy of the incoming signal to harvest energy, which in turn, is employed to meet the power requirements of the IT and IR sections. The transmission/reflection coefficient for the $i$-th element of the IT/IR section is $\varphi_{X,i}=\sqrt{\Upsilon_X} e^{j\theta_{X,i}} \:\:i=1,\dots,N_X,$
where $X \in \{ t,r \}$, $\Upsilon_X \in [0,\Upsilon_{\max}]$ is the amplification factor and $\theta_{X,i} \in [0,2\pi)$ is the phase shift. Based on the channel state information (CSI), the IT/IR section intelligently amplifies the incoming signal and provides the desired phase shifts by varying the diagonal transmission and reflection matrix
\vspace{-1mm}
\begin{equation}  \label{phidef}
\Phi_X= {\rm diag} \left( \varphi_{X,1},\dots,\varphi_{X,N_X} \right), \quad X \in \{ t,r \}.
\end{equation}

\subsection{Channel Model}  \label{chmodel}
We assume that the wireless links suffer from both large-scale path loss effects and small-scale fading. Specifically, the received power at the MF-RIS is attenuated by a factor of $C_0d_{sr}^{-\alpha_{sr}}$, where $C_0$ is the path loss at a reference distance, $d_{sr}$ is the distance between the Tx and the MF-RIS, and $\alpha_{sr}$ is the corresponding path loss exponent \cite{zeris}. Moreover, we consider a frequency-selective channel between the Tx and the MF-RIS with $L_{sr}$ independent paths. The received signal, at an arbitrary $n$-th element of the MF-RIS $(n=1,\dots,N)$ is
\vspace{-1mm}
\begin{equation}    \label{yris}
y_{p,q,n}=\sqrt{P_{\rm t}C_0d_{sr}^{-\alpha_{sr}}}\sum\limits_{l=1}^{L_{sr}} h_{l,n}s_{p,q-\tau_l}+w_{n},
\end{equation}
where $P_{\rm t}$ is the transmission power, $h_{l,n}$ and $s_{p,q-\tau_l}$ denote the complex channel coefficient and the delayed transmitted signal corresponding to the $l$-th path, respectively, and $w_{n}$ is the additive white Gaussian noise (AWGN) with zero mean and variance $\frac{N_0}{2}$. We assume that $|h_{l,n}|=\alpha_{l,n}$ is a Rayleigh random variable \cite{chaoswipt4} with $\mathbb{E}\{\alpha_{l,n}^2\}=\Omega_{\alpha,l,n}$, $\sum_{l=1}^{L_{sr}} \Omega_{\alpha,l,n}=1$ $\forall$ $n$, and identical channel statistics  across all the elements, i.e., $\Omega_{\alpha,l,1}=\Omega_{\alpha,l,2}=\dots=\Omega_{\alpha,l,N}=\Omega_{\alpha,l}$ $l=1,\dots,L_{sr}$.

Similarly, with $X \in \{ t,r \}$, the received power at $U_X$ is again attenuated by $C_0d_{rd_X}^{-\alpha_{rd_X}}$, with $d_{rd_X}$ being the distance between the MF-RIS and $U_X$ and $\alpha_{rd_X}$ is the corresponding path loss exponent. Lastly, here also, we model the channel between the MF-RIS and $U_X$ as a $L_{rd_X}$ tap channel with complex coefficient $g_{X,n,l}$ $(n=1,\dots,N_X \:\: l=1,\dots,L_{rd_X})$, where $|g_{X,n,l}|=\beta_{X,n,l}$ is Rayleigh distributed with $\mathbb{E}\{\beta_{X,n,l}^2\}=\Omega_{\beta,X,n,l}$, $\sum_{l=1}^{L_{rd_X}} \Omega_{\beta,X,n,l}=1$ $\forall$ $n$, and identical channel statistics across all MF-RIS elements.

\section{MF-RIS-aided Chaotic System Design}
In this section, the SR-DCSK frame structure is considered in the context of our MF-RIS-aided network\footnote{This work can also be generalized in the context of the newly proposed stacked intelligent metasurfaces, which is essentially composed of stacked multi-layer reconfigurable surfaces \cite{r3r1}.}. Note that, power transfer occurs here at the EH section, while information transfer takes place through the IT and IR section, respectively.

\subsection{Proposed System Design}
Practical scenarios suggest that the largest path delay is much smaller as compared to the reference length $\phi$ \cite{chaoswipt4,twc}. Hence, by using \eqref{yris}, the signal incident to the $n$-th MF-RIS patch $(n=1,\dots,N)$ can be expressed as
\vspace{-1mm}
\begin{equation}  \label{sigrisdef}
y_{p,q,n} \approx \sqrt{P_{\rm t}C_0d_{sr}^{-\alpha_{sr}}}\sum\limits_{l=1}^{L_{sr}} h_{l,n}s_{p,q}+w_{n}.
\end{equation}
Therefore, the resulting signal vector at the MF-RIS is
\vspace{-1mm}
\begin{equation}  \label{rsigris}
\boldsymbol{y_{p,q}}= \sqrt{P_{\rm t}C_0d_{sr}^{-\alpha_{sr}}}\boldsymbol{\rm h}s_{p,q}+\boldsymbol{\rm w},
\end{equation}
where $\boldsymbol{y_{p,q}}\!=\!\left[ y_{p,q,1}, y_{p,q,2}, \cdots, y_{p,q,N} \right]^T,$ $\boldsymbol{\rm h}=\left[ \sum_{l=1}^{L_{sr}} h_{l,1},\sum_{l=1}^{L_{sr}} h_{l,2}, \cdots, \sum_{l=1}^{L_{sr}} h_{l,N} \right]^T\!$, and $\boldsymbol{\rm w}=\left[ w_{1}, w_{2}, \right.$ $\left. \cdots, w_{N} \right]^T \!$.
Since a fraction of $\boldsymbol{y_{p,q}}$ is used for EH, IT, and IR, respectively, we rewrite \eqref{rsigris} as $\boldsymbol{y_{p,q}}= \left[ \left( \boldsymbol{y_{p,q}^{\rm EH}} \right)^T \left( \boldsymbol{y_{p,q}^{\rm IT}} \right)^T \left( \boldsymbol{y_{p,q}^{\rm IR}} \right)^T \right]^T,$
where $\boldsymbol{y_{p,q}^{\rm EH}}=$ $\!\Big[ y_{p,q,1},
\cdots, y_{p,q,N_h} \Big]^T, \boldsymbol{y_{p,q}^{\rm IT}}\!=\!\Big[ y_{p,q,N_h+1},\cdots,$ $ y_{p,q,N_h+N_t} \Big]^T$, and $\boldsymbol{y_{p,q}^{\rm IR}}\!=\!\Big[ y_{p,q,N_h+N_t+1},
\cdots, y_{p,q,N} \Big]^T$ denote the received signal vector at the EH, IT, and IR section, respectively.

As $\boldsymbol{y_{p,q}^{\rm IT}}$ is transmitted to $U_t$ by using the phase shift matrix $\Phi_t$, the received signal at $U_t$ is $y_{p,q}^t = \sqrt{C_0d_{rd_t}^{-\alpha_{rd_t}}}\boldsymbol{g_t}^T \Phi_t \boldsymbol{y_{p,q}^{\rm IT}}+w_t,$
where $\boldsymbol{g_t}=$ $\left[
\sum_{k=1}^{L_{rd_t}} g_{t,1,k}  
\sum_{k=1}^{L_{rd_t}} g_{t,2,k}
\cdots,
\sum_{k=1}^{L_{rd_t}} g_{t,N_t,k}
\right]^T$ and $w_t$ is the AWGN at $U_t$ with zero mean and variance $\frac{N_{t,0}}{2}$. By using \eqref{rsigris} and defining $\delta_t=\sqrt{P_{\rm t}C_0^2 d_{sr}^{-\alpha_{sr}} d_{rd_t}^{-\alpha_{rd_t}}\Upsilon_t}$ and $\Psi_t=$ $\sqrt{C_0 d_{rd_t}^{-\alpha_{rd_t}}\Upsilon_t}$, we rewrite the above expression to obtain
\vspace{-1mm}
\begin{align}
y_{p,q}^t& =\underbrace{\delta_t\sum\limits_{l=1}^{L_{sr}}\sum\limits_{k=1}^{L_{rd_t}} \sum\limits_{n=1}^{N_t} e^{j\theta_{t,n}}h_{l,N_h+n}g_{t,n,k}s_{p,q}}_{{\rm Desired \: signal \: at} \: U_t} \nonumber \\
& +\underbrace{\Psi_t\sum\limits_{k=1}^{L_{rd_t}} \sum\limits_{n=1}^{N_t}e^{j\theta_{t,n}}g_{t,n,k}w_{N_h+n}}_{\rm MF-RIS \: noise}+w_t.
\end{align}
Moreover, DCSK being a noncoherent modulation technique, CSI is unavailable at the MF-RIS. This results in an imperfect phase correction and as a result, the received signal at $U_t$ is
\vspace{-2mm}
\begin{align}  \label{trcv}
y_{p,q}^t& =\delta_t\sum\limits_{l=1}^{L_{sr}}\sum\limits_{k=1}^{L_{rd_t}} \sum\limits_{n=1}^{N_t} e^{j\theta_{t,e,n}}\alpha_{l,N_h+n}\beta_{t,n,k}s_{p,q} \nonumber \\
& +\Psi_t\sum\limits_{k=1}^{L_{rd_t}} \sum\limits_{n=1}^{N_t}e^{j\theta_{t,e,n}^w}\beta_{t,n,k}w_{N_h+n}+w_t,
\end{align}
where we define $\theta_{t,e,n}=\theta_{t,n}+\angle h_{l,N_h+n}+\angle g_{t,n,k}$ and $\theta_{t,e,n}^w=\theta_{t,n}+\angle g_{t,n,k}$ $\forall$ $k,l,n$. Thereafter, we define $\delta_r=\sqrt{P_{\rm t}C_0^2 d_{sr}^{-\alpha_{sr}} d_{rd_r}^{-\alpha_{rd_r}}\Upsilon_r},\Psi_r=\sqrt{C_0 d_{rd_r}^{-\alpha_{rd_r}}\Upsilon_r}$ and follow a similar procedure to obtain the received signal at $U_r$
\vspace{-1mm}
\begin{align}  \label{rrcv}
y_{p,q}^r& =\delta_r\sum\limits_{l=1}^{L_{sr}}\sum\limits_{k=1}^{L_{rd_t}} \sum\limits_{n=1}^{N_r} e^{j\theta_{r,e,n}}\alpha_{l,N_h+N_t+n}\beta_{r,n,k}s_{p,q} \nonumber \\
& +\Psi_r\sum\limits_{k=1}^{L_{rd_r}} \sum\limits_{n=1}^{N_r}e^{j\theta_{r,e,n}^w}\beta_{r,n,k}w_{N_h+N_t+n}+w_r,
\end{align}
where we have $\theta_{r,e,n}=\theta_{r,n}+\angle h_{l,N_h+N_t+n}+\angle g_{r,n,k}$ and $\theta_{r,e,n}^w=\theta_{r,n}+\angle g_{r,n,k}$ $\forall$ $k,l,n$, and $w_r$ is the AWGN at $U_r$ with zero mean and variance $\frac{N_{r,0}}{2}$.

\vspace{-1mm}
\begin{figure*}
\begin{align}  \label{dec}
\lambda_p^{(X)}&= \Re \left( T_c \sum\limits_{b=1}^{\zeta}\sum\limits_{z=0}^{\phi-1} \Bigg( \delta_X\sum\limits_{l=1}^{L_{sr}}\sum\limits_{k=1}^{L_{rd_X}} \sum\limits_{n=1}^{N_X} e^{j\theta_{e,X,n}}\alpha_{l,N_h+n}\beta_{X,n,k}x_{p,z}d_p + \Psi_X \sum\limits_{k=1}^{L_{rd_X}} \sum\limits_{n=1}^{N_X} e^{j\theta^w_{e,X,n}}\beta_{X,n,k}w_{N_h+n} +w_{X,b,z+\phi} \Bigg) \right. \!\!\! \nonumber \\
& \times \left.\Bigg( \delta_X\sum\limits_{l=1}^{L_{sr}}\sum\limits_{k=1}^{L_{rd_X}} \sum\limits_{n=1}^{N_X} e^{j\theta_{e,X,n}}\alpha_{l,N_h+n}\beta_{X,n,k}x_{p,z} + \Psi_X \sum\limits_{k=1}^{L_{rd_X}} \sum\limits_{n=1}^{N_X} e^{j\theta^w_{e,X,n}}\beta_{X,n,k}w_{N_h+n} +w_{X,z} \Bigg)^* \right)\!\!. 
\end{align}
\hrule
\vspace{-6mm}
\end{figure*}

\subsection{Information and Power Transfer}
Both $U_t$ and $U_r$ recover the chaotic component from the transmitted frame to perform $\zeta$ partial correlations over each block of $\phi$ samples. By considering the aspect of low cross-correlation of two chaotic sequences \cite{tcom}, the decision metric $\lambda_p^{(X)}$ $X \in \{ t,r \}$ corresponding to the $p$-th transmission interval, is obtained from \eqref{trcv} and \eqref{rrcv}, respectively, as expressed in \eqref{dec}, where $^*$ denotes the complex conjugate. Thereafter, $\lambda_p^{(X)}$ is compared with a threshold and the actual transmitted data is recovered from the received signal.

At the EH section of the MF-RIS, each element is connected to an individual EH unit consisting of a diode followed by a low pass filter \cite{comb}. To enhance the EH performance, we employ a $\phi+\beta$ bit analog correlator \cite{jstsp} prior to each EH unit.  An analog correlator is a low complexity yet very effective signal integrator, which simply sums the received signal over a certain time interval. The resultant output signal, corresponding to the $p$-th transmission interval, at the $n$-th element of the EH section is given by
\vspace{-1mm}
\begin{equation}
y_{p,n}^{\rm C}=\sum_{q=1}^{\phi+\beta}y_{p,q,n}\overset{(a)}{=} \sqrt{P_{\rm t}C_0d_{sr}^{-\alpha_{sr}}}\left(\sum\limits_{l=1}^{L_{sr}} h_{l,n}\right) \left( \sum\limits_{q=1}^{\phi+\beta} s_{p,q} \right),
\end{equation} 
where $(a)$ follows from \eqref{sigrisdef}, by assuming that the noise is too small to be harvested. Therefore, based on the nonlinearities of the EH unit, the resulting output power is $P_{\rm EH}=\sum_{n=1}^{N_h}\frac{v_{n,out}^2}{R_L}$, where $R_L$ is the load resistance and $v_{n,out}=\eta_1 \mathbb{E} \{ |y_{p,n}^{\rm C}|^2 \} + \eta_2 \mathbb{E} \{ |y_{p,n}^{\rm C}|^4 \},$ with $\eta_1,\eta_2$ being system parameters \cite{comb}.

\section{Waveform Design Investigating Transmit-Reflect-Harvest Trade-off}
In this section, based on the acceptable application specific BER requirement at both $U_t$ and $U_r$, we characterize the transmit/reflect/harvest trade-off at the MF-RIS. Accordingly, we propose the best $(N_t,N_r,N_h)$ combination and its impact on the SR-DCSK-based transmit waveform design.

By considering a chip duration $T_c$, the transmitted bit energy is given by $E_b=P_{\rm t}T_c \left( \beta+\phi \right) \mathbb{E} \{ x^2 \}$,
where $x$ is the chaotic chip. Accordingly, the system BER performance is characterized by the following proposition.
\begin{prop}  \label{theo2}
The system BER of the MF-RIS is given by
\vspace{-1mm}
\begin{align}  \label{theober}
{\rm BER}_X &=\frac{1}{2}\int\limits_0^{\infty}\int\limits_0^{\infty} {\rm erfc} \left( \left[ \frac{(\beta+\phi)^2}{\gamma_{X,0}\Lambda_{X,1}\beta} \left( \frac{1}{\phi}+\frac{1}{2\gamma_{X,0}\Lambda_{X,1}} \right.\right.\right. \nonumber \\
& \!\!\!\!\!\!\!\!\!\!\!\!\left.\left. \left.+\frac{\Psi_X^2 \Lambda_{X,2}}{2\gamma_{X,0}\Lambda_{X,1}} \left(\frac{N_0}{N_{X,0}}\right) \left( \frac{\beta+\phi}{\phi} + \frac{4\beta\gamma_{X,0}\Lambda_{X,1 }}{\phi \left( \beta+\phi \right)} \right)\right)\right]^{-\frac{1}{2}}   \right) \nonumber \\
& \!\!\!\!\!\!\!\!\!\!\times f(\Lambda_{X,1})f(\Lambda_{X,2}) d\Lambda_{X,1}d\Lambda_{X,2},
\end{align}
where $X \in \{ t,r \},\gamma_{X,0}=  \frac{E_{\rm b}C_0^2 d_{sr}^{-\alpha_{sr}}d_{rd_X}^{-\alpha_{rd_X}}\Upsilon_X}{N_{t,0}}$, $\Lambda_{X,1} =\sum_{l=1}^{L_{sr}}\!\sum_{k=1}^{L_{rd_X}}\!\! \Big|\!\sum_{n=1}^{N_X}\!e^{j\theta_{X,e,n}}\alpha_{l,N_h+n}\beta_{X,n,k}\Big|^2$, and $\Lambda_{X,2} =\sum_{k=1}^{L_{rd_X}}\!\! \Big|\! \sum_{n=1}^{N_X}\! e^{j\theta_{X,e,n}^w}\beta_{X,n,k}\Big|^2.$ Here, ${\rm erfc}(\cdot)$ is the complementary error function and $f(\Lambda_{X,1})$ and $f(\Lambda_{X,2})$ is the probability density function of $\Lambda_{X,1}$ and $\Lambda_{X,2}$, respectively.
\end{prop}
\begin{proof}
See the Appendix.
\end{proof}
The probability density function of the form  $\Big|\!\sum_{n=1}^{N_X}\!e^{j\theta_{X,e,n}}\alpha_{l,N_h+n}\beta_{X,n,k}\Big|$ is investigated in \cite{dnaka} for a Nakagami-$m$ fading scenario. Hence, we can evaluate  $f(\Lambda_{X,1})$ and $f(\Lambda_{X,2})$ by using $m=1$ in \cite[Proposition 1]{dnaka} followed by the standard technique of transformation of random variables. We observe that, ${\rm BER}_X$ is a joint function of the channel characteristics, $\Phi_X$ (as defined in \eqref{phidef}), $\beta$, and $\phi$. Note that, $\Upsilon_X=1$ results in negligible impact of the noise at the MF-RIS on ${\rm BER}_X$. Taking this into account, and by using $\Upsilon_t=0,\Upsilon_r=1$, the  self-sustainable RIS discussed in \cite{tcom} is demonstrated to be a special case of the MF-RIS. This can also be observed from the fact that by replacing $\Upsilon_t=0,\Upsilon_r=1$ in \eqref{theober} and ignoring the MF-RIS generated noise at $U_X$, we obtain the BER as stated in \cite[Theorem 1]{tcom}. Moreover, $\Upsilon_X$ also depends on the distance between the MF-RIS and $U_X$. Therefore, in general, we have $\Upsilon_t \neq \Upsilon_r$ and hence, we individually obtain the BER for $U_t$ and $U_r$. Furthermore, \eqref{theober} also demonstrates the impact of the noise generated at the MF-RIS on the BER performance. Specifically, based on the nature of ${\rm erfc}(x)$ for $x \geq 0$, we conclude that, for a given set of system parameters, an increasing $N_0$ translates to a deteriorating BER performance.

Next, we look into the performance of the EH section of the MF-RIS, which consists of $N_h$ elements. Accordingly, by using the Nakagami-$m$ parameter as unity and $\Gamma(1.5)=\frac{\sqrt{\pi}}{2}$ in  \cite[Theorem 3]{tcom}, the total harvested power is evaluated as
\vspace{-1mm}
\begin{align}  \label{ehf}
P_{\rm EH}&=\frac{N_h}{R_L} \left( \nu_1\chi_1\phi \left( 1+\zeta^2 \right) \right. \nonumber \\
& \left. + 9\nu_2\chi_2\phi \left( 1+6\zeta^2+\zeta^4 \right) \left( 2\phi-1 \right) \right)^2,
\end{align}
where $\nu_1=\eta_1P_{\rm t}C_0d_{sr}^{-\alpha_{sr}},\nu_2=\eta_2 \left( P_{\rm t}C_0d_{sr}^{-\alpha_{sr}}\right)^2$,
\vspace{-2mm}
\begin{equation*}
\chi_1=\frac{1}{4} \left( 2+\pi \sum_{\mathclap{\substack{l_1,l_2=1\\l_1 \neq l_2}}}^{L_{sr}} \sqrt{\Omega_{\alpha,l_1}\Omega_{\alpha,l_2}} \right),
\end{equation*}
and
\vspace{-2mm}
\begin{align*}
\chi_2&=\sum\limits_{k_1+k_2+\dots+k_{L_{sr}}=4}\! \frac{1}{k_1! \: k_2! \: \dots \: k_{L_{sr}}!}\! \prod\limits_{l=1}^{L_{sr}} \Gamma \left( 1+\frac{k_l}{2} \right) \Omega_{\alpha,l}^{\frac{k_l}{2}}.
\end{align*}
Note that, for a SR-DCSK frame of length $\beta+\phi$, the total energy consumption of the MF-RIS is
\vspace{-1mm}
\begin{equation} \label{ereq}
E_{\rm req}=T_c(\beta+\phi)E_{\rm net},
\end{equation}
\begin{align*} 
{\rm where} \:\:E_{\rm net}=& N_h P_{\rm conv}+ \left( N_t+N_r \right) \left( P_{\rm C}+P_{\rm DC} \right) \nonumber \\
& \!\!\!\!\!\!\!\!\!\!\!\!\!\!\!\!\!\!\!\!\!\!\!\!\!\!\!\!\!\!\!\!\!\!+ \xi \left( P_{\rm t}C_0d_{sr}^{-\alpha_{sr}} \left( \Upsilon_t \! \sum\limits_{l=1}^{L_{sr}}\!\sum\limits_{n=1}^{N_t}\!\alpha_{l,N_h+n}^2 \right.\right. \nonumber \\
& \!\!\!\!\!\!\!\!\!\!\!\!\!\!\!\!\!\!\!\!\!\!\!\!\!\!\!\!\!\!\!\!\!\!\!\!\left. \left.+ \Upsilon_r \! \sum\limits_{l=1}^{L_{sr}}\!\sum\limits_{n=1}^{N_r}\! \alpha_{l,N_h+N_t+n}^2 \right)\! + \! \frac{L_{sr}}{2}\left( \Upsilon_tN_tN_{t,0}+\Upsilon_rN_rN_{r,0} \right) \right)\!.
\end{align*}
Here, $P_{\rm conv},P_{\rm C},$ and $P_{\rm DC}$ denote the power consumed by the EH unit, power consumed by each phase shifter, and the DC baising power consumed by the amplifier, respectively and $\xi$ is the inverse of the amplifier efficiency \cite{mfris1}. To ensure self-sustainability of the MF-RIS, we must ensure $P_{\rm EH} \geq E_{\rm req}$.

From \eqref{theober}, we can state that for a given set of system parameters, ${\rm BER}_X$ is a function of $N_X$. Accordingly, from \eqref{ehf}, we obtain a lower bound on $N_h$ as
\vspace{-1mm}
\begin{align}  \label{nhbound}
N_h & \geq N_h^{\min} \nonumber \\
&= \frac{E_{\rm req}R_L}{\left( \nu_1\chi_1\phi \left( 1+\zeta^2 \right)+ 9\nu_2\chi_2\phi \left( 1+6\zeta^2+\zeta^4 \right) \left( 2\phi-1 \right) \right)^2}.
\end{align}
We observe that, with the other parameters remaining constant, $E_{\rm req}$ increases with $\Upsilon_t$ and $\Upsilon_r$. Therefore, even with identical choice of the $(N_t,N_r)$ combination, different $\Upsilon_t$ and $\Upsilon_r$ yield different $N_h^{\min}$. Moreover, the above bound inherently assumes $N_h^{\min} \leq N-(N_t+N_r)$, which may not always hold. In such a scenario, an appropriate choice of transmit waveform parameters for a fixed $(N_t,N_r,N_h)$ combination becomes extremely crucial. Therefore, for a fixed $(N_t,N_r,N_h)$ combination, we investigate the role of transmit waveform parameters on the BER-harvested power trade-off. Note that, to the best of our knowledge, no such investigation exists  in the literature for a chaotic MF-RIS-aided system. Hence, based on the definition of the `success rate' ${\rm SR}_X=1-{\rm BER}_X$ \cite{twc}, we propose the generalized ${\rm SR}-P_{\rm EH}$ region definition as
\vspace{-1mm}
\begin{align}  \label{region}
& \mathcal{C}_{{\rm SR}-P_{\rm EH}} \left( \phi: \phi \geq \max \left\lbrace \phi_{t,0}, \phi_{r,0} \right\rbrace \right) \nonumber \\
&= \left\lbrace \left( {\rm SR}_t,{\rm SR}_r,P_{\rm EH} \right): {\rm SR}_X \geq {\rm SR}_{X,0}, P_{\rm EH} \geq E_{\rm req} \right\rbrace,
\end{align}
where ${\rm SR}_{X,0}$ is the application-specific minimum acceptable success rate for $X \in \{t,r\}$ and the criteria of $P_{\rm EH} \geq E_{\rm req}$ guarantees the self-sustainability condition of the MF-RIS. With $\Upsilon_r=1,\Upsilon_t=0,N_t=0,$ and $N_h+N_r=N,$ the above characterization coincides with the one proposed in \cite{tcom}.

\section{Numerical Results and Conclusion}
We consider a transmission power $P_{\rm t}=30$ dBm, path loss at one meter distance $C_0=10^{-3.53}$, and the noise power at the MF-RIS, $U_t$, and $U_r$ is $-90$ dBm. We have the path loss exponent $\alpha_{sr}=\alpha_{rd_t}=\alpha_{rd_r}=3$ and a two-tap Rayleigh fading wireless channel with $\Omega_{\alpha,1}=\Omega_{\beta,t,1}=\Omega_{\beta,r,1}=0.8$, and $\Omega_{\alpha,2}=\Omega_{\beta,t,2}=\Omega_{\beta,r,2}=0.2$. The parameters for the considered non-linear EH model are: $\eta_1=0.9207 \times 10^3,\eta_2=0.0052 \times 10^9,$ and $R_L=5000$ $\Omega$ \cite{comb}.

\begin{figure*}[!t]
 \begin{subfigure}[b]{.316\textwidth}
    \centering
    \includegraphics[width=\linewidth]{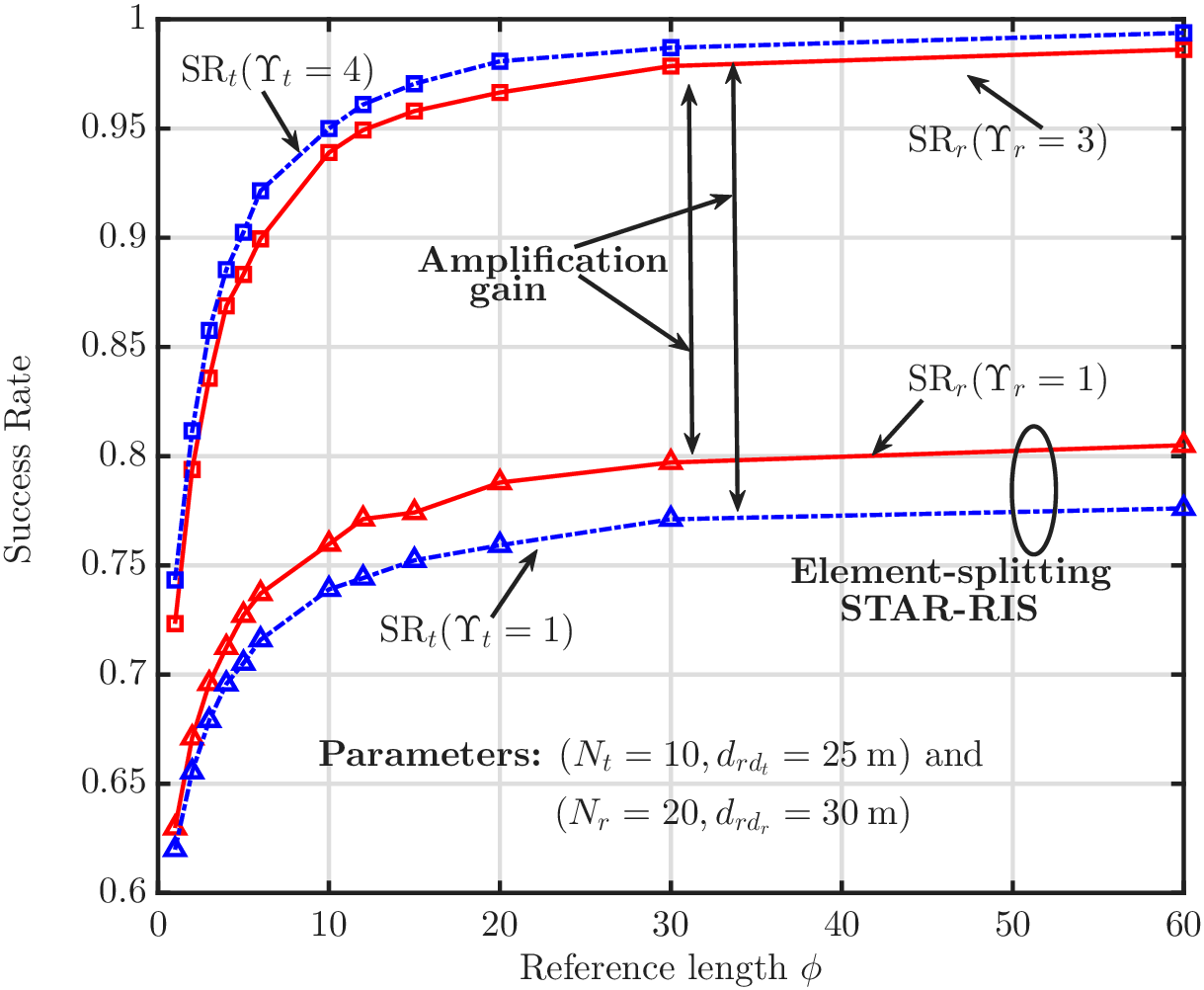}
    \vspace{-2mm}
    \caption{}
    \vspace{-2mm}
    \label{fig:res1}
\end{subfigure}
\begin{subfigure}[b]{.3\textwidth}
    \centering
    \includegraphics[width=\linewidth]{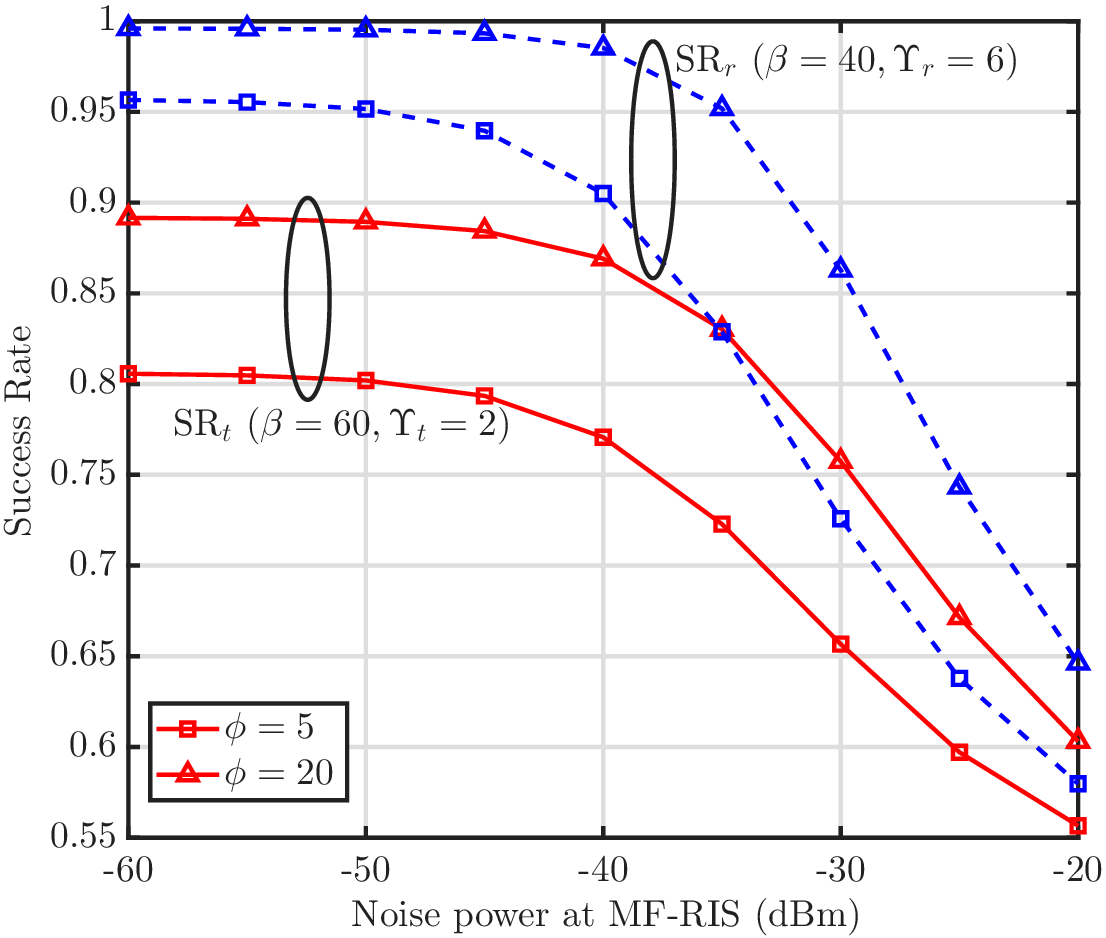}
    \vspace{-2mm}
    \caption{}
    \vspace{-2mm}
    \label{fig:res1a}
\end{subfigure}
\begin{subfigure}[b]{.34\textwidth}
    \centering
    \includegraphics[width=\linewidth]{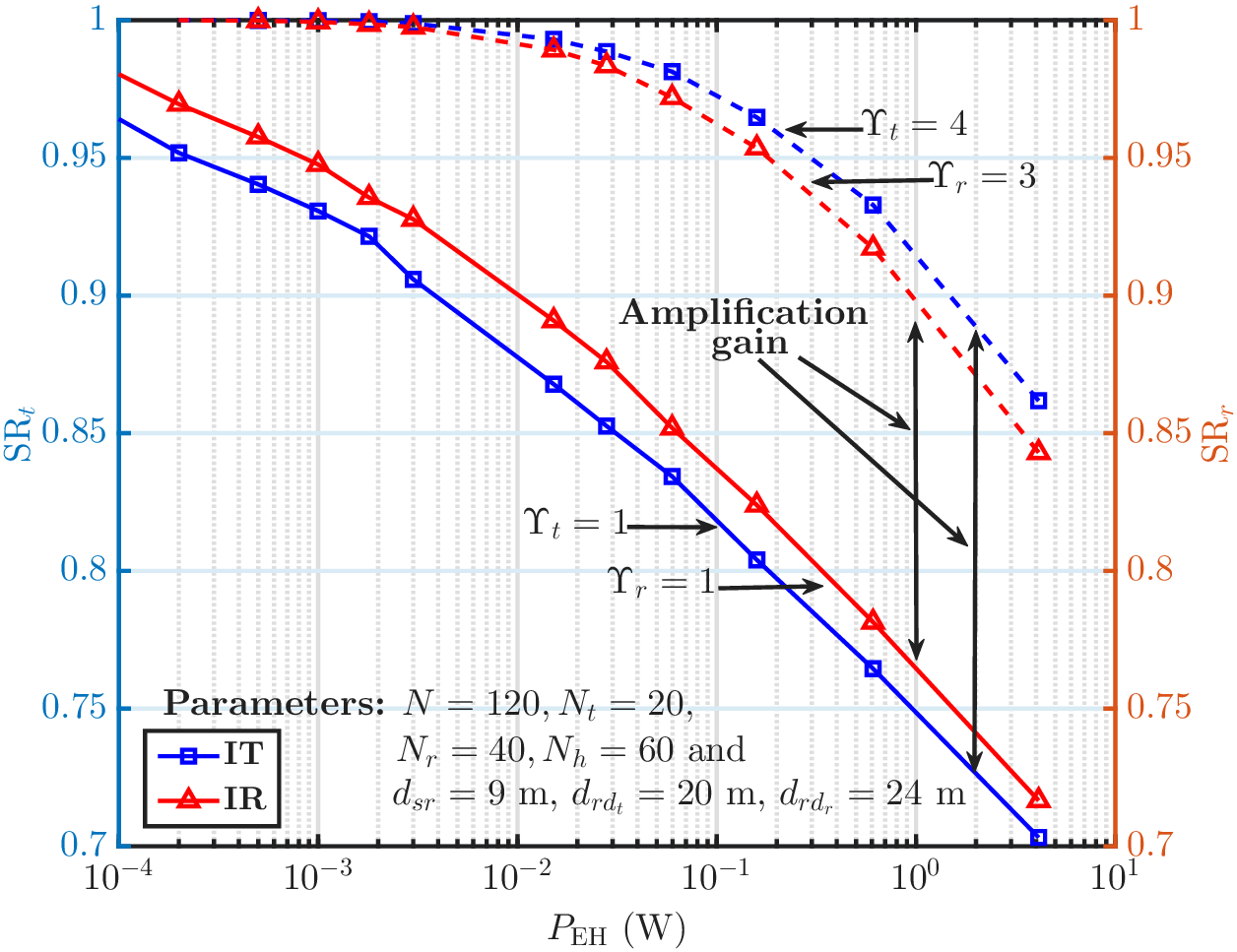}
    \vspace{-2mm}
     \caption{}
    \vspace{-4mm}
    \label{fig:res2}
\end{subfigure}
\caption{\footnotesize  (a) Impact of $\phi$ on ${\rm SR}$, (b) Impact of noise power at MF-RIS on ${\rm SR}$, and (c) Impact of system parameters on the ${\rm SR}-P_{\rm EH}$ region.}
\vspace{-2mm}
\end{figure*}

Fig. \ref{fig:res1} shows the impact of $\phi$ and the amplification factors $\Upsilon_X$ on the SR performance. By considering $\beta=60$ and the system parameters as stated in the figure, we vary $\phi$ to investigate its impact on ${\rm SR}_X$ for various $\Upsilon_X$. We observe that irrespective of $\Upsilon_X$, ${\rm SR}_X$ monotonically increases with $\phi$; the rate of increase in ${\rm SR}_X$ is initially linear but it saturates gradually. Also, we note a significant improvement in the ${\rm SR}_X$ performance with higher $\Upsilon_X$; for example, observe the performance gap at $\phi=30$ between ${\rm SR}_r$ for $\Upsilon_r=1$ and $\Upsilon_r=3$. However, this enhancement in ${\rm SR}$ performance comes at the cost of a higher $E_{\rm req}$. In other words, \eqref{ereq} shows that a higher $\Upsilon_X$ results in a higher $E_{\rm req}$, i.e., more energy is required at the MF-RIS to attain self-sustainability. Lastly, with $\Upsilon_t=\Upsilon_r=1$, the BER performance merges with that of the element splitting STAR-RIS \cite{sris}. Fig. \ref{fig:res1a} illustrates that for identical system parameters, {\rm SR} decreases with increasing MF-RIS noise, which is intuitive and also, can be observed from Proposition \ref{theo2}.

Fig. \ref{fig:res2} illustrates the proposed characterization with the considered system parameters and also, spreading factor $\beta=60$. Specifically, for this set of system parameters, we vary the reference length $\phi$ in $[1,\beta]$ such that $\zeta=\frac{\beta}{\phi} \in \mathbb{Z}$ and investigate its combined impact on ${\rm SR}_t,{\rm SR}_r,$ and $P_{\rm EH}$, respectively. We observe that, for a given $(N_t,N_r,N_h)$ combination, having $\Upsilon_t>1$ and $\Upsilon_r>1$ results in significant enhancement of the proposed generalized ${\rm SR}-P_{\rm EH}$ region. Moreover, the figure demonstrates the inevitable role of transmit waveform design on the system performance. Conventional DCSK, i.e., $\zeta=1$ (leftmost points in all the plots) results in the best IT and IR performance but a poor EH performance, which implies a very high $N_h^{\min}$ in \eqref{nhbound}. On the contrary, $\zeta=\beta$ results in a completely opposite scenario, i.e., worst IT and IR performance but excellent EH performance. Hence, DCSK is not always the best option and depending on the application specific acceptable BER, we decide on $\phi$ for a given $\beta$.

Therefore, in this work, we proposed a DCSK-based MF-RIS-aided noncoherent SWIPT framework. While the IT and IR section of the MF-RIS serve the users in the transmission and reflection region, respectively, the EH section harvests energy from the incident signal to meet their energy requirement. Accordingly, we derived the analytical expression for the system BER and characterized the harvested energy-BER trade-off. We also investigated the impact of this trade-off on the problem of transmit waveform design.

\appendix

To evaluate ${\rm BER}_X$, we need to obtain the mean and variance of $\lambda_p^{(X)}$ from \eqref{dec} \cite{r1r1}. Accordingly, by using $\beta=\zeta\phi$ and the definition $E_b=P_{\rm t}T_c \left( \beta+\phi \right) \mathbb{E} \{ x^2 \}$, we get
\vspace{-1mm}
\begin{align}  \label{meanvar}
\mathbb{E}\{ \lambda_p^{(X)} \}&=\frac{\beta\gamma_{X,0}}{\left( \beta+\phi \right)}N_{X,0} d_p\Lambda_{X,1} \quad {\rm and} \nonumber \\
{\rm var}\{ \lambda_p^{(X)} \}&= \frac{\beta N_{X,0}^2}{2} \left(\Lambda_{X,1} \frac{\gamma_{X,0} \left( \zeta+1 \right)}{\beta+\phi}  +  \frac{1}{2}  \right) \nonumber \\
& \!\!\!\!\!\!\!\!\!\!\!\!\!\!\!\!\!\!\! + \frac{\beta \Psi_X^2 N_0N_{X,0}}{4}\Lambda_{X,2} \left( \left( \zeta+1 \right)+ \frac{4\zeta\gamma_{X,0}}{\left( \beta+\phi \right)} \Lambda_{X,1} \right),
\end{align}
where $\mathbb{E}\{\cdot\}$ denotes the expectation operatorn. Note that, both $\lambda_p|(d_p=+1)$ and $\lambda_p|(d_p=-1)$ are sum of a significantly large number of random variables. Therefore, by using the Gaussian approximation \cite{chaoswipt4} and assuming equally probable transmission of $d_p=\pm 1$, the system IT/IR BER based on the channel conditions is obtained as
\vspace{-1mm}
\begin{align} \label{ber}
&{\rm BER}_X \left( \Lambda_{X,1},\Lambda_{X,2} \right) \nonumber \\
&=\!\frac{1}{2}\mathbb{P}\! \left\lbrace \! \lambda_p^{(X)}<1 | d_p=+1 \!\right\rbrace \! + \! \frac{1}{2}\mathbb{P}\! \left\lbrace \!\lambda_p^{(X)}>1 | d_p=-1 \!\right\rbrace \nonumber \\
&=\!\frac{1}{2}{\rm erfc} \left( \left[\frac{2 \:{\rm var}\left\lbrace \! \lambda_p^{(X)}| d_p=+1 \!\right\rbrace}{\mathbb{E}^2\left\lbrace \! \lambda_p^{(X)}| d_p=+1 \!\right\rbrace } \right]^{-\frac{1}{2}}   \right).
\end{align}
By replacing \eqref{meanvar} in \eqref{ber} and followed by generalization over the considered channel model, we obtain
\begin{equation*}
{\rm BER}_X\!\!=\!\!\!\!\int\limits_0^{\infty} \!\!\int\limits_0^{\infty}\!\! {\rm BER}_X \! \left( \Lambda_{X,1},\Lambda_{X,2} \right)\!f(\Lambda_{X,1})\!f(\Lambda_{X,2}) d\Lambda_{X,1}d\Lambda_{X,2}.
\end{equation*}

\bibliographystyle{IEEEtran}
\bibliography{refs}
\end{document}